\documentclass[pra,aps,twocolumn,showpacs,superscriptaddress]{revtex4-1}
\usepackage{graphicx}
\usepackage{multirow}
\usepackage{amsmath}
\usepackage{amssymb}
\usepackage{amsthm}
\usepackage{eucal}
\usepackage{bm}
\usepackage{mathrsfs}
\usepackage{latexsym}
\usepackage{float}
\begin{document}
\title{ Is nonclassicality-breaking the same thing as entanglement-breaking?}
\author{J. Solomon Ivan}
\email{solomonivan@iist.ac.in}
\affiliation{Indian Institute of Space Science and Technology, Valiamala, 
Trivandrum 695 547.}
\author{Krishna Kumar Sabapathy}
\email{kkumar@imsc.res.in}
\affiliation{Optics \& Quantum Information Group\\ The Institute of Mathematical
  Sciences, C.I.T Campus, Tharamani, Chennai 600 113.}
\author{R. Simon}
\email{simon@imsc.res.in}
\affiliation{Optics \& Quantum Information Group\\ The Institute of Mathematical
  Sciences, C.I.T Campus, Tharamani, Chennai 600 113.}

\begin{abstract}
{ 
Nonclassicality and entanglement are notions fundamental to quantum information processes involving continuous variable systems. That these two notions are intimately related has been intuitively appreciated for quite some time. An aspect of considerable interest is the behaviour of these attributes of a state under the action of a noisy channel. Inspired by the notion of entanglement-breaking channels, we define the concept of nonclassicality-breaking channels in a natural manner. We show that the notion of nonclassicality-breaking is essentially equivalent---in a clearly defined sense of the phrase `essentially'---to the notion of entanglement-breaking, as far as bosonic Gaussian channels are concerned. This is notwithstanding the fact that the very notion of entanglement-breaking requires reference to a bipartite system, whereas the definition of nonclassicality-breaking makes no such reference. Our analysis rests on our classification of channels into nonclassicality-based, as against entanglement-based, types of canonical forms. Our result takes ones intuitive understanding of the close relationship between nonclassicality and entanglement a step closer. }
\end{abstract}
\pacs{03.67.Mn, 42.50.-p, 03.65.Yz, 42.50.Dv, 03.67.-a}
\maketitle

\section{Introduction}
Two notions that have been particularly well explored in the context of
quantum information of continuous variable states are 
{\em nonclassicality}\,\cite{sudarshan63} and
{\em entanglement}\,\cite{schrodinger35}. 
The `older' notion of entanglement has become one of renewed interest 
in recent decades for its central role and applications in (potential as well as demonstrated)
quantum information processes\,\cite{ref2}, while the concept of nonclassicality, which emerges
directly from the {\em diagonal representation}\,\cite{sudarshan63}
had already been well explored in the quantum optical context\,\cite{ref1},
even before the emergence of the present quantum information era.
A fundamental distinction between these two notions may be noted\,: 
{\em While nonclassicality can be defined even for states of a
single mode of radiation, the very notion of entanglement requires two or more parties}.
Nevertheless, it turns out that the two notions are not entirely independent of one another; they are rather
intimately related\,\cite{simon00, asboth05, solomon11}. 
In fact, nonclassicality is a prerequisite
for entanglement\,\cite{solomon11}. Since a nonclassical bipartite state
whose nonclassicality  can be removed by local unitaries could not be entangled,
one can assert, at least in an intuitive sense, that {\em entanglement is nonlocal
nonclassicality}.

An important aspect in the study of nonclassicality and entanglement
is in regard of their evolution under the action of a channel. 
A noisy channel acting on a state can degrade its nonclassical
features\,\cite{ref3}. 
Similarly, entanglement can be degraded by channels acting  
locally on the constituent parties or modes\,\cite{ref4, ref4b,horodecki03, holevo08}.    
In fact, there are channels that render every bipartite state separable by
acting on just one of the parties\,\cite{horodecki03,holevo08,shirkov05}.
Such channels are said to be {\em entanglement-breaking}. 
We may recall that a channel $\Gamma$ is a linear completely positive trace-preserving
map that takes a state $\hat{\rho}_{\rm a}$ of a system $A$
to state $\hat{\rho}_{\rm a'}$ of system $A'$. That is, 
$\hat{\rho}_{\rm a'} = \Gamma(\hat{\rho}_{\rm a})\geq 0,\,\,{\rm Tr}(\hat{\rho}_{\rm a'})=1$ 
for every $\hat{\rho}_{\rm a} \geq 0$, 
${\rm Tr}(\hat{\rho}_{\rm a})=1$. Further, 
$\hat{\rho}_{\rm a' e} = \Gamma \otimes {1\!\!1}_{e}\,(\hat{\rho}_{\rm ae})$
is a physical state (i.e., unit-trace positive operator) for every input state $\hat{\rho}_{\rm ae}$ of the 
extended composite system $A+E$, with the environment $E$ assumed to be arbitrary\,\cite{ref5}\,: this is the notion of complete positivity (CP).

In the present work we address the following issue\,: {\em which channels
possess the property of ridding every input state of its nonclassicality?}
Inspired by the notion of entanglement-breaking channels, we may call
such channels {\em nonclassicality-breaking channels}. The close
connection between nonclassicality and entanglement alluded to
earlier raises a related second issue\,: {\em what is the connection,
if any, between entanglement-breaking channels and nonclassicality-breaking
channels?} To appreciate the nontriviality of the second issue, it suffices
to simply note that the very definition of entanglement-breaking refers
to bipartite states whereas the notion of nonclassicality-breaking
makes no such reference. In this paper we show that both these issues
can be completely answered in the case of bosonic Gaussian channels\,:
nonclassicality-breaking channels are enumerated, and it is shown that the set of all
nonclassicality-breaking channels is essentially the same as the set of all entanglement-breaking channels. 

We hasten to clarify the caveat `essentially'. Suppose a channel $\Gamma$ is nonclassicality-breaking as well as entanglement-breaking, and let us follow the action of this channel with a local unitary ${\cal U}$. The composite ${\cal U}\,\Gamma$ is clearly entanglement-breaking. But local unitaries can create nonclassicality, and so ${\cal U}\,\Gamma$ need not be nonclassicality-breaking. We say $\Gamma$ is {\em essentially nonclassicality-breaking} 
if there exists a fixed unitary ${\cal U}$ dependent on $\Gamma$ but independent of the input state 
on which $\Gamma$ acts, so that ${\cal U}\,\Gamma$ is nonclassicality-breaking. We may stress that this definition is not vacuous, for  given a collection of states {\em it is generically the case that there is no single unitary which would render the entire set nonclassical}. [This is not necessarily a property of the collection\,: given a nonclassical mixed state $\rho$, it is possibly not guaranteed that there exists an unitary ${\cal U}$ such that $\hat{\rho}^{\,'} = {\cal U}\, \hat{\rho} \,{\cal U}^{\dagger}$ is classical.] It is thus reasonable to declare the set of entanglement-breaking  channels to be the same as the set of nonclassicality-breaking channels if at all the two sets indeed turn out to be the same, modulo this `obvious' caveat or provision. 

Gaussian channels are 
physical processes that map Gaussian states to
Gaussian states. They are generalization of symplectic (metaplectic) unitaries, generated by Hamiltonians quadratic in the mode operators, which too map Gaussian (pure and mixed) states into Gaussian states\,\cite{simon87,simon88,dutta94,arvind95}. 
To realise a Gaussian channel, the state of the system is coupled to 
a Gaussian state of an
ancilla system of modes,  evolved jointly using a symplectic unitary, and then the ancilla modes 
are discarded. Gaussian channels have played an important role in
quantum information processing with continuous variable states, 
and this has lead to their systematic 
analysis\,\cite{eisert07, holevo07, werner02,gaussch, caruso08, wolf07, ref6}. 
Single-mode Gaussian channels were classified 
in\,\cite{holevo07}, and their canonical forms were enumerated. 
Their operator sum representation
was obtained in\,\cite{gaussch}. Multi-mode Gaussian channels and 
their canonical forms were studied in\,\cite{caruso08}.


The outline of the presentation is as follows. 
Section II contains a brief discussion on the concept of
$s$-ordered quasi-probabilities and their corresponding $s$-ordered 
characteristic functions. This is done in anticipation of
its use as the principal tool in our entire analysis. 
The diagonal representation ($s=1$) and the important
notion that arises from it---the classicality-nonclassicality divide---are noted, 
the classicality-nonclassicality
divide leading, inspired by the notion of entanglement-breaking channels, to a natural definition of 
the notion of nonclassicality-breaking channels.
We briefly discuss Gaussian states 
and bosonic Gaussian channels in Section III, including a brief consideration
of 
entanglement-breaking Gaussian channels.
In Section IV we present a complete classification of 
single-mode Gaussian channels into 
{\em classicality-based} canonical forms. There are three different canonical forms, and these are distinct from the entanglement-based
 canonical  forms obtained by Holevo and 
collaborators\,\cite{holevo07,werner02},
 the notion of 
nonclassicality-breaking having a more restricted invariance than the notion of entanglement-breaking. Necessary and sufficient condition on the channel parameters, in order that the channel breaks nonclassicality of every input state, is derived in Section V for each of the three nonclassicality-based canonical forms.  In Section VI we present a comparative analysis
of  nonclassicality-breaking and entanglement-breaking channels. 
The paper concludes with some final remarks in Section VII.

\section{nonclassicality-breaking channels}
A state of a quantum mechanical system specified by  
density operator $\hat{\rho}$ can be faithfully described 
by any  member of the one-parameter family of $s$-ordered quasi-probability distributions
or, equivalently, by the corresponding $s$-ordered characteristic function\,\cite{cahill691}. 
For a single mode of radiation field with mode
operators $\hat{a}$ and $\hat{a}^{\dagger}$ satisfying the 
commutation relation $[ \hat{a}, \hat{a}^{\dagger}]={1\!\!1}$,
the $s$-ordered  characteristic function associated with state $\hat{\rho}$
is defined as\,\cite{cahill691}
\begin{eqnarray}
\chi_{s}(\xi; \rho) = \exp \left[\,\frac{s}{2} |\xi|^{2}\,\right]\, {\rm 
Tr}(\hat{\rho} D(\xi)), \,\,\,\,\, -1 \leq s \leq 1.
\label{quas1}
\end{eqnarray}
Here $\xi=(\xi_{1}+ i\xi_{2})/\sqrt{2} \in {\cal C}$, 
$D(\xi) = \exp (\xi \hat{a}^{\dagger}-\xi^{*}\hat{a})$ is the phase space displacement operator, and
$s$ is the order parameter. The particular cases $s= 1, 0, -1$ correspond, 
respectively, to normal-ordering $N$, Weyl or symmetric-ordering $W$, 
and antinormal-ordering $A$ of the mode operators.

By performing Fourier transformation on the $s$-ordered characteristic 
function $\chi_{s}(\xi; \rho)$, we obtain
\begin{eqnarray}
\!\!\!W_{s}(\alpha;\rho)\! =  
\frac{1}{\pi}\int \!\!{\rm exp}
[(\alpha {\xi}^{*} -{\alpha}^{*} \xi)\,]\,
\chi_{s}(\xi; \rho) d^2 \xi,
\label{quas2}
\end{eqnarray}
the corresponding $s$-ordered quasi-probability,
where ${\alpha}$ stands for the classical (c-number) phase space variable\,: 
${\alpha} =(q,\,p)= (q + i p)/\sqrt{2}
\in {\cal C}$.  The particular cases $s= -1, 0, 1$ correspond, 
respectively, to the better known $Q$ function, the Wigner function, 
and the diagonal `weight' function (also called the $P$ function).

Except the $Q$ function 
$Q(\alpha)=\langle \alpha | \hat{\rho} | \alpha \rangle$, which 
by definition is manifestly pointwise nonnegative over the complex plane ${\cal C}$, 
all other $s$-ordered quasi-probabilities assume negative values 
for some $\alpha$, at least for some states. That is, the $Q$ function
alone is a genuine probability distribution; but every genuine 
probability distribution over ${\cal C}$ is not a $Q$ function.

It is clear from\,(\ref{quas1}) that the characteristic functions of 
a state $\hat{\rho}$ for two different values $s_1$, $s_2$ of the `order parameter' 
$s$ are related as 
\begin{eqnarray}
\chi_{s_1}(\xi ; \rho)=
{\rm exp}\left(-(s_2-s_1){|\xi|}^2\right) \chi_{s_2}(\xi; \rho).
\label{quas2a}
\end{eqnarray}
Performing Fourier transformation, we see that the respective $s$-ordered 
quasi-probabilities (with $s_2 > s_1$) are related through a 
Gaussian convolution\,\cite{cahill691}.

Any density operator $\hat{\rho}$ representing some state of a single mode of 
radiation field can always be expanded as
\begin{eqnarray}
\hat{\rho} = \int\frac{d^2\alpha}{\pi}\, {\phi}_{\rho} (\alpha) | \alpha \rangle \langle \alpha |,
\label{quas3}
\end{eqnarray}
where ${\phi}_{\rho} (\alpha)= W_{1}(\alpha;\rho)$ is the diagonal `weight' function, 
$| \alpha \rangle$ being the
coherent state. This {\em diagonal representation} is made possible because of the 
over-completeness property
of the coherent state `basis'\,\cite{sudarshan63}. The diagonal representation\,(\ref{quas3})
enables the evaluation, {\em in a classical-looking manner},
of ensemble averages of normal-ordered operators, and this is
important from the experimental point of view\,\cite{wolf-book}.


An important notion that arises from the diagonal representation 
is the {\em classicality-nonclassicality divide}. If $\phi_{\rho}(\alpha)$
associated with density operator $\hat{\rho}$ is pointwise nonnegative
over ${\cal C}$, then the state is a convex sum, or ensemble,
of coherent states. Since coherent states are the most elementary 
of all quantum mechanical states exhibiting classical behaviour, any state
that can be written as a convex sum of these elementary classical states
is deemed classical. We have,
\begin{eqnarray}
{\phi}_{\rho} (\alpha) \geq 0 \,\,\,\,{\rm for}\,\,\,{\rm all} \,\,\,\alpha
\in {\cal C}\,\Leftrightarrow\, \hat{\rho}\,\,{\rm is}\,\,{\rm classical}.
\label{quas4}
\end{eqnarray}      
Any state which cannot be so written is declared to be  nonclassical. Fock states $| n\rangle\langle n|$,
whose diagonal weight function
$\phi_{|n \rangle\langle n|}(\alpha)$ is the 
${\rm n^{th}}$ derivative of the delta function, are examples of nonclassical states. [All the above considerations generalize from one mode to $n$-modes in a painless manner, with $\alpha, \,\xi \in\, {\cal R}^{2n} \sim {\cal C}^n$.]

This classicality-nonclassicality divide leads to the following natural definition, inspired by the notion of entanglement-breaking\,:
\\

\noindent
{\em Definition}\,: A channel $\Gamma$ is said to be {\em nonclassicality-breaking}
if and only if the output state $\hat{\rho}_{\rm out}= \Gamma (\hat{\rho}_{\rm in})$ is classical
{\em for every} input state $\hat{\rho}_{\rm in}$, i.e., if and only if the diagonal
function of every output state is a genuine probability distribution.
\\

\section{Gaussian states and Gaussian channels}
A state $\hat{\rho}$ is said to be Gaussian if its $s$-ordered quasi-probability
or, equivalently, its $s$-ordered characteristic function is Gaussian.
And without loss of generality we may assume it to be a zero mean state.
The symmetric or Weyl-ordered characteristic 
function\,($s=0$) of a Gaussian state then has the form\,\cite{simon87,simon88} 
\begin{eqnarray}
\chi_{W}(\xi;\rho) = \exp\left[-\frac{\xi^T V \xi}{2} \right],
\label{ncb1}
\end{eqnarray} 
where $V$ is its variance matrix.
$V$ is real, symmetric, positive definite, and specifies the Gaussian state completely;
and $V$ necessarily obeys the uncertainty principle\,\cite{dutta94} 
\begin{eqnarray}
V + i \Sigma \geq 0, \,\,\,\,\Sigma= i\sigma_2 \oplus i\sigma_2 \oplus \cdots
\oplus i \sigma_2,
\label{ncb2}
\end{eqnarray}
where $\sigma_2$ is the antisymmetric Pauli matrix.

A Gaussian channel maps every Gaussian state to a Gaussian state.  
The action of a Gaussian channel thus 
manifests simply as a linear transformation on the variance matrix $V$. Under
the action of a Gaussian channel described by $(X, Y)$\,\cite{werner02},
\begin{eqnarray}
V \rightarrow V' = X^T V X +Y,
\label{ncb3}
\end{eqnarray} 
$Y$ being symmetric positive semidefinite.

For arbitrary input state with symmetric-ordered characteristic function 
$\chi_{W}(\xi;\rho)$, we have 
\begin{eqnarray}
\chi_{W}^{\rm in}(\xi;\rho) \rightarrow \chi^{\rm out}_{W}(\xi;\rho) = 
\chi_{W}(X \xi;\rho) \exp\left[-\frac{\xi^T
    Y \xi}{2} \right].
\label{ncb4}
\end{eqnarray}
If a single-mode Gaussian channel $(X,Y)$ acts on the $A$-mode 
of a two-mode squeezed
vacuum state $|\psi_{r}\rangle ={\rm sech}\, r \sum_{k=0}^{\infty}({\rm tanh}\, r)^{k}|k,\,k\rangle$, 
whose variance matrix equals $c_{2r}{1\!\!1}_{4 \times 4} + s_{2r}\sigma_1 \otimes \sigma_3$,
the result is a two-mode mixed Gaussian state specified by variance matrix
\begin{eqnarray}
V_{\rm out}(r)=\left(
\begin{matrix}
c_{2r} (X^TX) +Y && s_{2r}(X^T \sigma_3) \\
s_{2r} (\sigma_3 X) && c_{2r} ({1\!\!1}_2)
\end{matrix} 
\right),
\label{ncb5}
\end{eqnarray}
where $c_{2r}=\cosh\, 2r$, $s_{2r}= \sinh\, 2r$,
and $\sigma_j$ are the Pauli matrices. 
It is clear that $V_{\rm out}(r)$ should obey the mandatory uncertainty principle
\begin{eqnarray}
V_{\rm out}(r) + i \Sigma \geq 0, \,\,\,\,{\rm for \,\,all}\,\,r,
\label{ncb5a}
\end{eqnarray}
where $\Sigma= i\sigma_{2} \oplus i\sigma_{2}$. In fact, this uncertainty principle 
is both a necessary and sufficient condition on $(X, Y)$ to be a Gaussian
channel, and it may be restated in the form\,\cite{wolf07}
\begin{eqnarray}
Y + i\sigma \geq iX \sigma X^T.
\label{ncb5b}
\end{eqnarray}
Since a noisy Gaussian channel preceded and/or succeeded by Gaussian unitary (noiseless)
channels
is a Gaussian channel, the double Gaussian unitary freedom can be used to bring
both $X$ and $Y$ to simpler canonical forms, as shown 
in\,\cite{holevo07}. The canonical forms so determined are useful, for instance,
in the study of entanglement-breaking Gaussian channels\,\cite{holevo08}. We recall that a channel
$\Gamma$ {\em acting on system $A$} is entanglement-breaking if the bipartite 
output state $(\Gamma \otimes {1\!\!1}_{\rm e})\,(\hat{\rho}_{\rm ae})$ is separable
for every input state $\hat{\rho}_{\rm ae}$, 
the ancilla system $E$ being arbitrary\,\cite{horodecki03}.

\section{Nonclassicality-based canonical forms for Gaussian channels}
The canonical forms for Gaussian channels have been described
by Holevo\,\cite{holevo07} and Werner and Holevo\,\cite{werner02}.
Let ${\cal S}$ denote an element of the symplectic group $Sp(2n,\,R)$ of 
linear canonical transformations and ${\cal U}({\cal S})$
the corresponding unitary (metaplectic) operator\,\cite{dutta94}. One often 
encounters situations wherein the aspects one is looking for
are invariant under local unitary operations, entanglement
being an example. In such cases a Gaussian channel $\Gamma$ is `equivalent'
to ${\cal U}({\cal S}^{'})\,\Gamma\,{\cal U}({\cal S})$, for
arbitrary symplectic group elements ${\cal S}$, ${\cal S}^{'}$.
The orbits or double cosets of equivalent channels in this sense are the ones
classified and enumerated by Holevo and 
collaborators\,\cite{holevo07, werner02}.

While the classification of Holevo and collaborators is entanglement-based, as just noted,
the notion of nonclassicality-breaking has {\em a more restricted invariance}.
A nonclassicality-breaking Gaussian channel $\Gamma$
preceded by any Gaussian unitary ${\cal U}({\cal S})$ is nonclassicality-breaking
if and only if $\Gamma$ itself is nonclassicality breaking.
In contradistinction, the nonclassicality breaking aspect of $\Gamma$
and ${\cal U}({\cal S})\,\Gamma$ [$\Gamma$ followed the Gaussian unitary
${\cal U}({\cal S})$] are not equivalent in general; they are equivalent
if and only if ${\cal S}$ is in the 
intersection $Sp(2n,\,R) \cap SO(2n,\,R) \sim U(n)$ of `symplectic phase
space rotations' or passive elements\,\cite{dutta94,arvind95}. In the single-mode case
this intersection is just the rotation group $SO(2) \subset Sp(2,\,R)$. We thus
need to classify single-mode Gaussian channels $\Gamma$ into
orbits or double cosets ${\cal U}({\cal R})\,\Gamma\,{\cal U}({\cal S})$,
${\cal S} \in Sp(2,\,R)$, ${\cal R} \in SO(2) \subset Sp(2, R)$.
Equivalently, we need to classify $(X, Y)$ into orbits 
$({\cal S}\,X\,{\cal R},\,{\cal R}^{T}\,Y\,{\cal R})$. It turns out that
there are three distinct canonical forms, and the type into which a given pair $(X,Y)$ belongs is  fully determined by ${\rm det}\,X$.
\vskip 0.1cm

\noindent
{\bf First canonical form\,: ${\rm {\bf det}}\, {\bf X > 0}$}.
A real $2 \times 2$ matrix $X$ with ${\rm det}\,X =\kappa^2 > 0$
is necessarily of the form $\kappa\,{\cal S}_{X}$ for some
${\cal S}_{X} \in Sp(2,\,R)$. Indeed we have ${\cal S}_{X}=({\rm det}\,X)^{-1/2}\,X$
Choose ${\cal R} \in SO(2)$ so as to
diagonalise $Y > 0$\,: ${\cal R}^{T}\,Y\,{\cal R}={\rm diag}(a, b)$.
With such an ${\cal R}$, the choice ${\cal S}= {\cal R}^{T}{\cal S}_{X}^{-1} \in Sp(2,\,R)$
takes $(X,\,Y)$ to the canonical form $(\kappa{1\!\!1}, \,{\rm diag}(a, b))$,
where $\kappa =\sqrt{{\rm det}\,X} > 0$, and $a,\,b$ are the eigenvalues of $Y$.
\vskip 0.1cm

\noindent
{\bf Second canonical form\,:} {${\rm{\bf det}} \, {\bf X <0}$}.
Again choose ${\cal R}$ so that ${\cal R}^{T} Y {\cal R}={\rm diag}(a, b)$.
Since ${\rm det}\, X < 0$, $X$ is necessarily of the form $\kappa\,{\cal S}_{X}\,\sigma_3$,
for some ${\cal S}_{X} \in Sp(2,\, R)$\,: ${\cal S}_{X}=({\rm det}\, X\sigma_3)^{-1/2}X\sigma_3$.
Since ${\cal R}\,\sigma_3\,{\cal R}=\sigma_3$ for
every ${\cal R} \in SO(2)$, it is clear that the choice ${\cal S}={\cal R}\,{\cal S}_{X}^{-1} \in Sp(2,\,R)$
takes $(X,\,Y)$ to the canonical form $(\kappa\,\sigma_3,\,{\rm diag}(a,b))$ in this
case, with $\kappa=\sqrt{{\rm det}\,X\sigma_3}$, and the parameters $a,\,b$ being the
eigenvalues of $Y$.
\vskip 0.1cm

\noindent
{\bf Third canonical form\,:} ${\rm{\bf det}}\,{\bf X=0}$. Let $\kappa$ be the singular
value of $X$; choose ${\cal R}', \,\,{\cal R} \in SO(2)$ such that 
${\cal R}' \,X \,{\cal R}={\rm diag}(\kappa, 0)$. It is clear that the choice
${\cal S}_{X}={\rm diag}(\kappa^{-1}, \kappa)\,{\cal R}'^{\,T} \in Sp(2,\,R)$ along
with ${\cal R} \in SO(2)$ takes $(X,\, Y)$ to the canonical form 
$({\rm diag}(1,0),\,Y_{0}={\cal R}^{T}\,Y\,{\cal R})$. $Y_{0}$ does not, of course,
assume any special form. But if $X=0$, then ${\cal R} \in SO(2)$  can be chosen so as to
diagonalise $Y$\,: in that case $Y_{0}=(a,b),\,\,a,\,b$ being the eigenvalues of $Y$.

\section{Nonclassicality-breaking Gaussian channels}
Having obtained the nonclassicality-based canonical forms of $(X,\,Y)$, 
we now derive the necessary and sufficient 
conditions for a single-mode Gaussian channel to be nonclassicality-breaking.
We do it for the three canonical forms in that order. 
\vskip 0.1cm

\noindent
{\bf First canonical form\,:}\,${\bf \bm{(}X,\,Y\bm{)}\bm{=}
\bm{\left(}\bm{\kappa}{1\!\!1},\,{\rm {\bf diag}}\bm{(}a,b\bm{)}\bm{\right)}}$.
There are three possibilities\,: $\kappa=1$, $\kappa< 1$, and $\kappa>1$. We begin with
$\kappa =1$; it happens that the analysis  extends quite easily to the other two cases
and, indeed, to the other two canonical forms as well. 
The action on the
normal-ordered characteristic function in this case is
\begin{align}
&\chi_{N}^{\rm in}(\xi_1, \,\xi_2; \rho) \rightarrow {\chi}_{N}^{\rm out}(\xi_1, \xi_2;\rho)\nonumber\\
&=
\exp\left[-\frac{a\,\xi_1^2}{2}-\frac{b\,\xi_2^2}{2}  \right] 
\chi_{N}^{\rm in}(\xi_1,\,\xi_2; \rho).
\label{nbcf1}
\end{align} 
[For clarity, we shall write the subscript of $\chi$ explicitly as $N$, $W$, or $A$ in place of 1, 0,  or -1]. It should be appreciated that {\em for this class of Gaussian channels} ($\kappa=1$) the above 
input-output relationship
holds even with the subscript $N$ replaced by $W$ or $A$ uniformly. Let us assume
$a,\,b > 1$ so that $a =1 + \epsilon_1$, $b= 1+ \epsilon_2$ with $\epsilon_1 , \, \epsilon_2 >0$.
The above input-output relationship can then be written in the form 
\begin{eqnarray*}
{\chi}_{N}^{\rm out}(\xi_1, \xi_2; \rho)=
\exp\left[-\frac{\epsilon_1\,\xi_1^2}{2}-\frac{\epsilon_2\,\xi_2^2}{2}  \right] 
\chi_{W}^{\rm in}(\xi_1,\,\xi_2; \rho).
\label{nbcf2}
\end{eqnarray*}
Note that the subscript of $\chi$ on the right hand side is now $W$ and not $N$.

Define $\lambda > 0$ through $\lambda^2 = \sqrt{\epsilon_2 /\epsilon_1}$, and rewrite
the input-output relationship in the suggestive form
\begin{align}
{\chi}_{N}^{\rm out}(\lambda\xi_1, \lambda^{-1}\xi_2; \rho)&= \exp\left[-\frac{1}{2}(\sqrt{\epsilon_1\epsilon_2}\,\xi_1^2-
\sqrt{\epsilon_1\epsilon_2}\,\xi_2^2)  \right]
 \nonumber\\
&~~~~~\times  \chi_{W}^{\rm in}(\lambda\xi_1,\,\lambda^{-1}\xi_2; \rho).
\label{nbcf3}
\end{align}
But $\chi_{W}^{\rm in}(\lambda\xi_1,\,\lambda^{-1}\xi_2; \rho)$ is simply the
Weyl-ordered or Wigner characteristic function of a (single-mode-)
squeezed version of $\hat{\rho}$, for every $\hat{\rho}$. If ${\cal U}_{\lambda}$
represents the unitary (metaplectic) operator that effects this squeezing
transformation specified by squeeze parameter $\lambda$, we have
\begin{eqnarray}
\chi_{W}^{\rm in}(\lambda\xi_1,\,\lambda^{-1}\xi_2; \rho)=
\chi_{W}^{\rm in}(\xi_1,\,\xi_2; {\cal U}_{\lambda}\,\rho\,{\cal U}_{\lambda}^{\dagger}),
\label{nbcf4}
\end{eqnarray}
so that the right hand side of the last input-output relationship, {\em in the
special case} $\epsilon_1 \epsilon_2 =1$, reads
\begin{eqnarray}
\chi_{W}^{\rm out}(\lambda\xi_1,\,\lambda^{-1}\xi_2; \rho)=
\chi_{A}^{\rm in}(\xi_1,\,\xi_2; {\cal U}_{\lambda}\,\rho\,{\cal U}_{\lambda}^{\dagger}).
\label{nbcf5}
\end{eqnarray}
This special case would transcribe, on Fourier transformation, to
\begin{eqnarray}
&&\phi^{\rm out}(\lambda \alpha_1,\,\lambda^{-1} \alpha_2; \rho)=
Q^{\rm in}(\alpha_1,\,\alpha_2; {\cal U}_{\lambda}\,\rho\,{\cal U}_{\lambda}^{\dagger})\nonumber \\
&&~~~~=\langle \alpha | {\cal U}_{\lambda}\,\hat{\rho}\,{\cal U}^{\dagger}_{\lambda}|\alpha\rangle
\geq 0,\,\, \forall\,\,\alpha,\,\,\forall\,\,\hat{\rho}.
\label{nbcf6}
\end{eqnarray}
That is, the output diagonal weight function evaluated at $(\lambda\alpha_1, \,\lambda^{-1}\alpha_2 )$ equals the input
$Q$-function evaluated at $(\alpha_1, \,\alpha_2)$, and hence is nonnegative for all $\alpha \in {\cal C}$. 
Thus the output state is classical
for every input, and hence the channel is nonclassicality-breaking. It is clear that if
$\epsilon_1 \epsilon_2 > 1$,
the further Gaussian convolution corresponding to the additional multiplicative factor
$\exp \left[-(\sqrt{\epsilon_1 \epsilon_2}-1) (\xi_{1}^{2}+\xi_{2}^{2})/2\right]$
in the output characteristic function will only render the output state even more strongly
classical. We have thus established this {\em sufficient condition}
\begin{eqnarray}
(a-1)(b-1)\geq 1,
\label{nbcf7}
\end{eqnarray} 
or, equivalently,
\begin{align}
\frac{1}{a}+ \frac{1}{b} \leq 1.
\label{nbcf7b}
\end{align}

Having derived a sufficient condition for nonclassicality-breaking, we 
derive a necessary condition by looking at the
signature of the output diagonal weight function {\em for a particular
input state} evaluated at {\em a particular phase space} point at the
output. Let the input be the Fock state $|1\rangle\langle1|$, the first excited
state of the oscillator. Fourier transforming the input-output relation (\ref{nbcf1}),
one readily computes the output diagonal weight function to be 
\begin{eqnarray}
&&{\phi}^{\rm out}(\alpha_1,\,\alpha_2; |1 \rangle\langle 1|) =   \frac{2}{\sqrt{ab}}\,
\exp\left[-\frac{2 \alpha_{1}^{2}}{a}
-\frac{2 \alpha_{2}^{2}}{b}\right]\,\nonumber\\
&& \hspace{2cm} \times \left(1+ \frac{4(\alpha_1 + \alpha_2)^2}{a^{2}} - \frac{1}{a} 
-\frac{1}{b}\right). 
\label{nbcf8}
\end{eqnarray}  
An obvious necessary condition for nonclassicality-breaking is that this
function should be nonnegative everywhere in  phase space. Nonnegativity at the single phase space point $\alpha =0$
gives the necessary condition ${1}/{a} + {1}/{b} \leq 1$ which is, perhaps
surprisingly, the same as the sufficiency condition established earlier! That is, {\em the  sufficient condition (\ref{nbcf7}) is also a necessary condition for nonclassicality-breaking}.
Saturation of this inequality corresponds to the boundary wherein the channel
is `just' nonclassicality-breaking. {\em The formal resemblance in this case
with the law of distances in respect of imaging by a thin convex lens} is unlikely to miss  the reader's attention.


The above proof for the particular case of classical noise channel $(\kappa=1)$ gets easily extended to noisy
beamsplitter (attenuator) channel $(\kappa < 1)$ and noisy
amplifier channel $(\kappa > 1)$. The action of the channel 
$(\kappa1\!\!1,\,{\rm diag}(a, b))$ on the
normal-ordered characteristic function follows from that on the Wigner
characteristic function given in\,(\ref{ncb4})\,: 
\begin{eqnarray}
{\chi}_{N}^{\rm out}(\xi; \rho)&=& \exp \left[ -\frac{\tilde{a}\, \xi_1^2}{2}
-\frac{\tilde{b}\,\xi_2^2}{2} \right]\chi_{N}^{\rm in}(\kappa\,\xi;\rho),  \nonumber \\
\tilde{a}&=& a+ \kappa^2 -1,\,\,\,\tilde{b}= b+\kappa^2 -1.  
\label{b2}
\end{eqnarray}
This may be rewritten in the suggestive form 
\begin{align}
{\chi}_{N}^{\rm out}(\kappa^{-1}\xi; \rho) = 
\exp \left[ -\frac{\tilde{a}\, \xi_1^2}{2\kappa^{2}}
  -\frac{\tilde{b}\, \xi_2^2}{2\kappa^{2}} \right]
\chi_{N}^{\rm in}(\xi; \rho).
\label{b3}
\end{align}  
With this we see that the right hand side of \eqref{b3} to be the same as right hand side of
\eqref{nbcf2} with
${\tilde{a}}/{\kappa^2}$, ${\tilde{b}}/{\kappa^2}$
replacing $a, \,b$. 
The case $\kappa \not= 1$ thus gets essentially reduced
to the case $\kappa=1$, the case of classical noise channel,
analysed in detail above. This leads to the following {\em necessary
and sufficient condition for nonclassicality-breaking} 
\begin{align}
&\frac{1}{a+\kappa^2-1} + \frac{1}{b+\kappa^2-1} \leq
\frac{1}{\kappa^2} \nonumber\\
&~ \Leftrightarrow ~ (a-1)(b-1) \geq  \kappa^4,
\label{b4}
\end{align}  
for all $\kappa>0$, thus completing our analysis of the first canonical form.
\vskip 0.1cm

\noindent
{\bf Second canonical form\,:}\,\,${\bf \bm{(}X,Y\bm{)}\bm{=}\bm{(}\bm{\kappa\,\sigma_3},
\,{\rm {\bf diag}}\bm{(}a,b\bm{)}\bm{)}}$.
The noisy phase conjugation channel with canonical form $(\kappa\,\sigma_3, \,{\rm diag}(a,b))$
acts on the normal-ordered characteristic
function in the following manner, as may be seen from its action on the Weyl-ordered
characteristic function (\ref{ncb4})\,:
\begin{eqnarray}
{\chi}_{N}^{\rm out}(\xi;\rho)= 
\exp \left[ -\frac{\tilde{a}\, \xi_1^2}{2}
  -\frac{\tilde{b}\, \xi_2^2}{2} \right]
\chi_{N}^{\rm in}(\kappa\,\sigma_3\,\xi; \rho),
\label{p2}
\end{eqnarray} 
with $\tilde{a}= a+ \kappa^2 -1,\,\,\,\tilde{b}= b+\kappa^2 -1$ again, and $\kappa \,\sigma_3\,\xi$
denoting the pair $(\kappa\,\xi_1, -\kappa\,\xi_2)$.
As in the case of the noisy amplifier/attenuator channel, we rewrite it in the
form
\begin{align}
{\chi}_{N}^{\rm out}(\kappa^{-1}\,\sigma_3\, \xi; \rho) = 
\exp \left[ -\frac{\tilde{a}\,\xi_1^2}{2\kappa^{2}}
  -\frac{\tilde{b}\,\xi_2^2}{2\kappa^{2}} \right]
\chi_{N}^{\rm in}(\xi; \rho), 
\label{p3}
\end{align}  
the right hand side of \eqref{p3} has the same form as (\ref{nbcf2}), leading to the
{\em necessary and sufficient nonclassicality-breaking condition}
\begin{align}
\frac{1}{\tilde{a}}+ \frac{1}{\tilde{b}} \leq \frac{1}{\kappa^2}
~~\Leftrightarrow~~
(a-1)(b-1) \geq  \kappa^4.
\label{p4}
\end{align}  

\noindent
{\bf Remark}\,: We note in passing that in exploiting the `similarity' of
Eqs.\,(\ref{b3}) and (\ref{p3}) with Eq.\,(\ref{nbcf2}),
we made use of the following two elementary facts\,: (1) An
invertible linear change of variables 
$[f(x) \rightarrow f(A\,x),\,\,{\rm det}\,A \not= 0]$
on a multivariable function $f(x)$ reflects as a
corresponding linear change of variables in its Fourier
transform\,; (2) A function $f(x)$ is pointwise
nonnegative if and only if $f(A\,x)$ is pointwise nonnegative
for every invertible $A$. In the case of (\ref{b3}), the linear change $A$
corresponds to uniform scaling, and in the case of (\ref{p3})
it corresponds to uniform scaling followed or preceded by 
mirror reflection. 
\vskip 0.1cm

\begin{table*}
\centering
\begin{tabular}{|c|c|c|c|}
\hline
Canonical form & Nonclassicality-breaking & Entanglement-breaking & Complete positivity \\
& condition & condition & condition \\
\hline
$(\kappa\,{1\!\!1},\, {\rm diag}(a,b))$  &   $(a-1)(b-1) \geq \kappa^4$ 
& $ab\geq(1+\kappa^2)^2$  & $ab\geq (1-\kappa^2)^2$\\
\hline
$(\kappa\,\sigma_3,\,{\rm diag}(a,b))$ & $(a-1)(b-1) \geq \kappa^4$ 
& $ab\geq(1+\kappa^2)^2$  & $ab\geq (1+\kappa^2)^2$\\
\hline
$({\rm diag}(1,0),\,Y)$, &$a,\, b \geq 1$, $a$, $b$ being & $ab\geq1$ & $a b \geq 1$ \\
& eigenvalues of $Y$ & &  \\
$({\rm diag}(0,0),\, {\rm diag}(a,b))$ &$a,\,b \geq 1$ & $ab\geq1$ & $ab \geq 1$ \\
\hline
\end{tabular}
\caption{A comparison of the nonclassicality-breaking condition, the entanglement-breaking
condition, and the complete positivity condition for the three canonical classes of channels.\label{table1}}
\end{table*}

\noindent
{\bf Third canonical form\,:} {\bf Singular $\bm{X}$}.
Unlike the previous two cases, it proves to be convenient to begin with the Weyl or symmetric-ordered 
characteristic function in this case of singular $X$\,:
\begin{eqnarray}
{\chi}_{W}^{\rm out}(\xi; \rho) = \exp \left[-\frac{1}{2}
  \xi^T \,Y_{0}\, \xi \right] \chi_{W}^{\rm in}(\xi_1 , 0; \rho).
\label{p5}
\end{eqnarray}
Since we are dealing with symmetric ordering, $\chi_{W}^{\rm in}(\xi_1 , 0; \rho)$
is the Fourier transform of the marginal distribution of the first quadrature
(`position' quadrature) variable. Let us assume that the input $\hat{\rho}$
is a (single-mode-) squeezed Gaussian pure state, squeezed in the position
(or first) quadrature. For arbitrarily large squeezing, the state approaches a position
eigenstate and the position quadrature marginal approaches the Dirac delta
function. That is $\chi_{W}^{\rm in}(\xi_1 , 0; \rho)$ approaches a constant.
Thus, the Gaussian $\exp \left[- (\xi^T \,Y_{0}\, \xi)/2 \right]$
is essentially the Weyl-characteristic function of the output state, and hence corresponds
to a classical state if and only if
\begin{eqnarray}
Y_{0} \geq 1\!\!1,\,\,\,{\rm or}\,\,\,a,\,\,b\geq 1,
\label{p6}
\end{eqnarray}   
$a,\,b$ being the eigenvalues of $Y$.

We have derived this as a {\em necessary condition for nonclassicality-breaking},
taking as input a highly squeezed state. It is clear that 
for any other input state the phase space distribution of the output state
will be a convolution of this Gaussian classical state with the position
quadrature marginal of the input state, rendering the output state more strongly
classical, and thus proving that the condition (\ref{p6}) is {\em also a sufficient
condition for nonclassicality-breaking}.  

In the special case in which $X=0$ identically, we have the following
input-output relation in place of (\ref{p5})\,:
\begin{eqnarray}
{\chi}_{W}^{\rm out}(\xi; \rho) = \exp \left[-\frac{1}{2}
  \xi^T \,Y\, \xi \right] \chi_{W}^{\rm in}(\xi=0; \rho).
\label{p7}
\end{eqnarray}
Since $\chi_{W}^{\rm in}(\xi=0; \rho)=1$ independent of $\hat{\rho}$, the output
is an input-independent {\em fixed state}, and $\exp \left[-\frac{1}{2}
  \xi^T \,Y\, \xi \right]$ is its Weyl-characteristic function. But we know that this
fixed output is a classical state if and only if $Y \geq 1\!\!1$. In other
words, {\em the condition for nonclassicality-breaking is the same for all
singular $X$, including vanishing $X$}. 

We conclude our analysis in this Section with the following, perhaps redundant,
remark\,: Since our canonical forms are nonclassicality-based,
rather than entanglement-based, if the nonclassicality-breaking property
applies for one member of an orbit or double coset, it applies to the
entire orbit.

\begin{figure*}
\centering
\scalebox{0.5}{\includegraphics{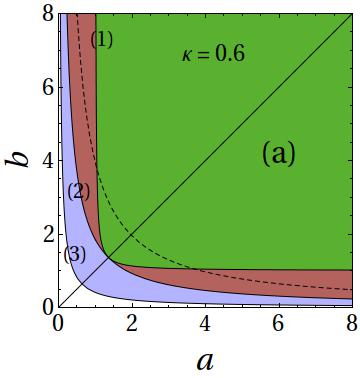}}~~ 
\scalebox{0.5}{\includegraphics{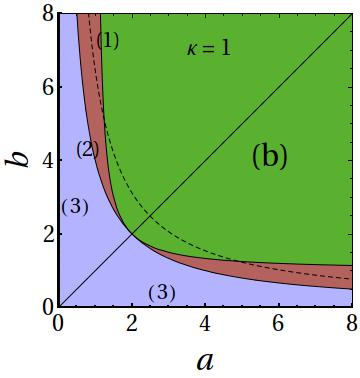}}\\
\scalebox{0.5}{\includegraphics{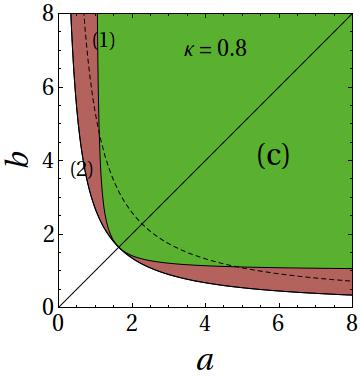}} ~~
\scalebox{0.5}{\includegraphics{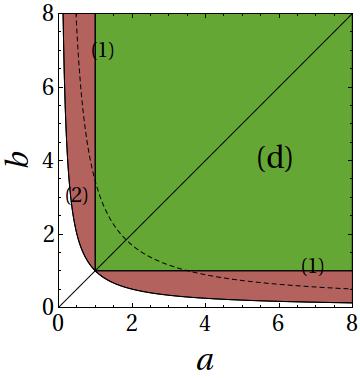}} 
\caption{
Showing a pictorial comparison of the nonclassicality-breaking condition, the entanglement-breaking condition, and the complete positivity condition in the channel parameter space $(a,b)$, for fixed ${\rm det}\, X$. Curves (1), (2), and (3) correspond to saturation of these conditions in that order. Curve (3) thus corresponds to quantum-limited channels. Frame (a) refers to the first canonical form $(\kappa1\!\!1,\,{\rm diag}(a,b))$,
frame (c) to the second canonical form $(\kappa\,\sigma_3,\,{\rm diag}(a,b))$,
and frame (d) to the third canonical form, singular $X$. Frame (b) 
refers to the limiting case $\kappa=1$, classical noise channel.   
In all the four frames, the region to the right of (above) curve (1) corresponds
to nonclassicality-breaking channels; the region to the right of (above) curve (2) corresponds
to entanglement-breaking channels; curve (3) depicts the CP condition, so the
region to the right of (above) it alone corresponds to physical channels. The region to the left (below) curve (3) is unphysical as channels. In frames
(c) and (d), curves (2) and (3) coincide. In frame (b), 
curve (3) of (a) reduces to the $a$ and $b$ axis shown in bold. In frames (a) and 
(c), curves (1) and (2) meet at the point $(1+\kappa^2, 1+\kappa^2)$, in frame (b) they meet  at $(2,2)$, and in frame (d) at $(1,1)$. 
The region between (2) and (3) corresponds to the set of channels which are not entanglement-breaking. That in frame (c) and (d) the two curves coincide proves that this set is vacuous for the second and third canonical forms. That in every frame the nonclassicality-breaking region is properly contained in the entanglement-breaking region proves that a nonclassicality-breaking channel is certainly an entanglement-breaking channel. The dotted curve in each frame indicates the orbit
of a generic entanglement-breaking Gaussian channel under the action of a local unitary squeezing after the channel action. That the orbit of every entanglement-breaking channel passes through the
nonclassicality-breaking region, proves that the nonclassicality in all the output
states of an entanglement-breaking channel can be removed by a fixed unitary squeezing, thus showing that every entanglement-breaking channel is `essentially' a nonclassicality-breaking channel. 
\label{fig}}
\end{figure*} 

\section{nonclassicality-breaking {\em vs} entanglement-breaking}
We are now fully equipped to explore the relationship between nonclassicality-breaking 
Gaussian channels and entanglement-breaking channels.
In the case of the first canonical form the nonclassicality-breaking condition reads $(a-1)(b-1)\geq \kappa^4$, the entanglement-breaking condition reads $ab \geq (1+\kappa^2)^2$, while the complete positivity condition reads $ab \geq (1-\kappa^2)^2$. These conditions are progressively weaker, indicating that the family of channels which meet these conditions are progressively larger. For the second canonical form the first two conditions have the same formal expression as the first canonical form, while the complete positivity condition has a more stringent form $ab \geq (1+\kappa^2)^2$. For the third and  final canonical form, the nonclassicality-breaking condition requires both $a$ and $b$ to be bounded from below by unity, whereas both the entanglement-breaking and  complete positivity conditions read $ab \geq 1$. Table\,\ref{table1} conveniently places these conditions side-by-side. In the case of  first canonical form, (first row of Table\,\ref{table1}), the complete positivity condition itself is vacuous for $\kappa=1$, the classical noise channels. 

This comparison is rendered pictorial in Fig.\,\ref{fig}, in the channel parameter plane $(a,b)$, for fixed values of ${\rm det} X$. Saturation of the nonclassicality-breaking condition, the entanglement-breaking condition, and the complete positivity condition are  marked $(1)$, $(2)$, and $(3)$ respectively in all the four frames. Frame (a) depicts the first canonical form for $\kappa=0.6$ (attenuator channel). The case of the amplifier channel takes a qualitatively similar form in this pictorial representation. As $\kappa \to 1$,  from below ($\kappa <1$) or above ($\kappa >1$), curve $(3)$ approaches the straight lines $a=0, \, b=0$ shown as solid lines in Frame (b) which depicts this limiting $\kappa=1$ case (the classical noise channel). Frame (c) corresponds to the second canonical form (phase conjugation channel) for $\kappa=0.8$ and Frame (d) to the third canonical form. It may be noticed that in Frames (c) and (d) the curves (2) and (3) merge, indicating and consistent with that fact that channels of the second and third canonical forms are aways entanglement-breaking.

It is clear that the nonclassicality-breaking condition is stronger than the entanglement-breaking condition. 
Thus, a nonclassicality-breaking
channel is necessarily entanglement-breaking\,: But there are channel 
parameter ranges wherein the channel is entanglement-breaking,
though not nonclassicality-breaking.
The dotted curves in Fig.\,\ref{fig} represent orbits of a generic entanglement-breaking channel $\Gamma$, fixed by the product $ab$ ($\kappa$ having been already fixed), when $\Gamma$ is followed up by a variable local unitary squeezing ${\cal U}(r)$. To see that the orbit of every entanglement-breaking channel passes through the nonclassicality-breaking region,
it suffices to note from Table\,\ref{table1} that the nonclassicality-breaking boundary has $a=1$, $b=1$ as asymptotes whereas the entanglement-breaking boundary has $a=0$, $b=0$ as the asymptotes.  
That is, for every entanglement-breaking channel there exists a particular
value of squeeze-parameter $r_{0}$, depending only on the channel
parameters and not on the input state, so that the entanglement-breaking
channel $\Gamma$ followed by unitary squeezing of extent $r_0$ always results
in a nonclassicality-breaking channel ${\cal U}(r_0)\,\Gamma$. It is in this precise sense 
that nonclassicality-breaking channels and entanglement-breaking channels
are essentially one and the same. 

Stated somewhat differently, if at all the output of an entanglement-breaking channel is nonclassical, the nonclassicality is of a `weak' kind in the following sense. Squeezing is not the only form of nonclassicality. Our result
not only says that the output of an entanglement-breaking 
channel could at the most have a squeezing-type nonclassicality,
it further says that the nonclassicality of {\em all} output states can be
removed by a {\em fixed} unitary squeezing transformation.

\section{Final Remarks}
We have explored the notion of nonclassicality-breaking and its relation to entanglement-breaking. We have shown that the two notions are effectively  equivalent in the context of bosonic Gaussian channels, even though at the level of definition the two notions are quite different, the latter requiring reference to a bipartite system. Our analysis shows that some nonclassicality could survive an entanglement-breaking channel, but this residual nonclassicality would be of a particular weaker kind. 

The close relationship between entanglement and nonclassicality has been studied by several authors in the past\,\cite{simon00,asboth05,solomon11,ref4b}. It would seem that our result brings this relationship another step closer.

Finally, we have presented details of the analysis only in the case of single-mode bosonic Gaussian channels. We believe the analysis is likely to generalize to the case of $n$-mode channels in a reasonably straight forward manner.

\end{document}